\documentclass[aip,amsmath,amssymb,reprint]{revtex4-2}

\usepackage{mathrsfs} %for Fourier F notation
\usepackage{color}
\usepackage{graphicx}% Include figure files
\usepackage{dcolumn}% Align table columns on decimal point
\usepackage{bm}% bold math
\usepackage{epstopdf} %converting to PDF

\usepackage[utf8]{inputenc}
\usepackage[T1]{fontenc} % Use modern font encodings
\usepackage{mathptmx}
\usepackage{etoolbox}

\usepackage{placeins} %keeps floats ‘in their place’, preventing them from floating past
\makeatletter
\def\@email#1#2{%
	\endgroup
	\patchcmd{\titleblock@produce}
	{\frontmatter@RRAPformat}
	{\frontmatter@RRAPformat{\produce@RRAP{*#1\href{mailto:#2}{#2}}}\frontmatter@RRAPformat}
	{}{}
}%
\makeatother
\usepackage{amstext}

\begin{document}
    
\title{Enhanced deep-tissue photoacoustics by using microcomposites made of radiofrequency metamaterials and soft polymers: Double- and triple-resonance phenomena}

\author{Ricardo Mart\'{\i}n Abraham-Ekeroth}
\affiliation{Instituto de F\'{\i}sica Arroyo Seco, IFAS (UNCPBA), Tandil, Argentina} 

\affiliation{CIFICEN (UNCPBA-CICPBA-CONICET), Grupo de Plasmas Densos, Pinto 399, 7000 Tandil, Argentina}
\email{mabraham@ifas.exa.unicen.edu.ar}

\date{\today}
    
%Keywords: Radiofrequency, Photoacoustics, Acoustic Resonances, Metamaterials, Ultrasound, Radio plasmonics, Spoof Plasmons
\begin{abstract}
   Photoacoustic imaging systems offer a platform with high resolution to explore body tissues, food, and artwork. On the other hand, plasmonics constitutes a source of resonant heating and thermal expansion to generate acoustic waves. However, its associated techniques are seriously limited to laser penetration and nonspecific hyperthermia in the sample.
   To address this issue, the present work adopts a paradigm shift in photoacoustics. By simulating microparticles made of  random composites, the calculated pressure can be made similar or superior to that calculated via plasmonic optoacoustics. The improvement is due to a phenomenon called double or triple resonance, which is the excitation of one or both electric and magnetic plasmons within radiofrequency range and the simultaneous excitation of the particle's acoustic mode. Given that electromagnetic pulses are restricted to nanosecond pulse widths and MHz frequencies, the proposed method overcomes the poor penetration in tissues and reduces thermal damage, thereby offering a noninvasive technique of theragnosis. Moreover, the resonant pressure obtained lasts longer  than  with conventional photoacoustic pressure, providing a central feature to enhance detection. To fully comprehend the multi-resonance framework, we develop a complete photoacoustic solution. The proposed approach could pave the way to thermoacoustic imaging and manipulation methods for sensitive materials and tissues with micrometer resolution.
\end{abstract}

\maketitle

\section{Introduction}

Photoacoustics dates back to 1880, when A. G. Bell heard a ``pure musical tone'' in a closed gas volume that had absorbed a modulated light beam. \cite{bell_production_1880-1} However,  interest in this field dropped off until relatively recently due to a lack of  technological advances. Interest resumed when the 1980s saw many experiments on photochemical reactions and sensitive gas detection.\cite{gusev_laser_1993} In the 1990s, with the advent of better technology, photoacoustic (PA) imaging became a promising technique for biomedical applications.\cite{xu_photoacoustic_2006,steinberg_photoacoustic_2019,choi_review_2019} Since then, various contrast agents have enabled image reconstruction from body tissues.\cite{mallidi_multiwavelength_2009,zhou_tutorial_2016,paltauf_progress_2020} However, it was not until the 2000s that  plasmonic nanoparticles made of noble metals were introduced in photoacoustics.\cite{wang_photoacoustic_2004,mantri_engineering_2020} As molecular dyes, plasmonic nanoparticles are advantageous with respect to previous contrast agents because they have high absorption efficiencies with a tunable spectral response due to their surface plasmon (SP) excitation in the optical range.\cite{kumarSimulationStudiesPhotoacoustic2018} Some nanoparticles have low cytotoxicity for biological tissues, are chemically inert, and are environmentally sensitive. They  can also be easily functionalized for different purposes.\cite{lee_simultaneous_2021} Thus, plasmonic-based techniques are still considered optimal for photothermal treatment of cancer cells because of the significant rise in temperature  that occurs as a result of optical absorption.\cite{baffou_thermo-plasmonics_2013,vines_gold_2019} 

However, high temperatures can be a major obstacle for noninvasive techniques. A significant rise in temperature can thermally damage  non-targeted tissues,\cite{zhao_recent_2021} even when using short laser pulses of a few nano- or femtoseconds.\cite{baffou_femtosecond-pulsed_2011} In addition, the effect leads to various undesirable phenomena like nano or microbubble formation or particle-shape modifications, both of which change the sensitivity to the environment and the spectral properties of the particles.\cite{lukianova-hleb_plasmonic_2010,prostPhotoacousticGenerationGold2015} The second main obstacle when using plasmonic-based methods is the relatively poor penetration in biological tissues because optical and NIR beams are required. In addition, the PA effect with nanoparticles strongly depends on the surface-to-volume ratio because the major component of the acoustic pressure generated is due to the embedding medium's layer that surrounds the particle.\cite{prostPhotoacousticGenerationGold2015} Thus, slight variations in the nanosources strongly affect the PA efficiency, causing instabilities.

This paper proposes a PA pressure-generation method that overcomes the obstacles related to optical plasmonics. Here, radioplasmonics \cite{abraham-ekeroth_radioplasmonics_2021,abraham-ekeroth_radioplasmonics_2021-4} is considered an extension of optical plasmonics to the radiofrequency (RF) range.\cite{liaoHorizontallyPolarizedOmnidirectional2020} RF surface plasmons can be excited in microparticles made of \textit{random} metamaterials [see recent reports of RF-metamaterials (RF-MMs)].\cite{tsutaokaLowFrequencyPlasmonic2013,houExperimentalRealizationTunable2015,shiPercolativeSilverAlumina2015,chengRadioFrequencyNegative2017,estevezTunableNegativePermittivity2019,sunTunableNegativePermittivity2019,tallmanEffectThermalLoading2020}. Briefly, RF-MMs are percolating composites made to work in the RF range. They can switch their electromagnetic (EM) character depending on the concentration of conductive fillers in their dielectric matrix around the percolation limit. Notably, the conductive fillers' concentration facilitates the electronic conduction within the composite (see Sec.\ref{sec:Theory}). The fillers' proximity manages free or hopping electrons; thus, driven electrons can move along short or long paths, make loops, or run into specific directions \cite{chengRadioFrequencyNegative2017}. As a result, RF-MMs can be modeled as capacitive or inductive by the number of gaps and loops the electrons find along their paths \cite{sunTunableNegativePermittivity2019,wangPermittivityTransitionPositive2020,wei_low-frequency_2021}. Thus, the effective permittivity and permeability of the composite vary. The constitutive functions determine the EM response of a MM so that a small piece of RF-MM can absorb highly long wavelengths. It occurs as if long effective electron paths were allowed to exist in RF-MMs, which seem to offer multiple rounds and random directions for the inner currents.

EM radiation at low energies allows for large penetration in tissues and resonant absorption due to plasmon excitation. Here, the simulation of microparticles made of RF-MMs and soft polymers shows that EM and acoustic resonances can match, which introduces a new concept into the discussion of photoacoutics: double and triple resonances. In the former, one electric or magnetic SP matches the monopolar acoustic mode of the particle. In the latter, both the electric and magnetic SPs match the particle's acoustic mode. The resulting PA pressure is greater and endures longer than is possible with optical fields and metal nanoparticles.

A complete analytical PA solution reveals a multi-resonance framework based on an arbitrary EM source. The leading example consists of a simulation of a Gaussian pulse exciting ``realistic'' metamaterials. The calculated PA pressure is enhanced with a negligible rise in temperature. A double-resonance effect was already predicted in Ref.~\cite{abraham-ekeroth_radioplasmonics_2021} by exciting the usual electric SP +  acoustic monopole. Here, the concept includes ``magnetic'' photoacoustics and triple-resonance phenomena.

The following section explains the electric and magnetic resonances that can occur in RF-MMs to define the proposed framework. 

The methodology used to simulate realistic composites is detailed. Because their EM, thermal, and mechanical properties are essential, suitable ``effective medium'' theories are applied when needed. They are described in the Supplemental Material  (SM) for the interested reader.

\section{Theoretical considerations \label{sec:Theory}}

\begin{figure*}%[H]
    \centering
    \includegraphics[width=160mm]{"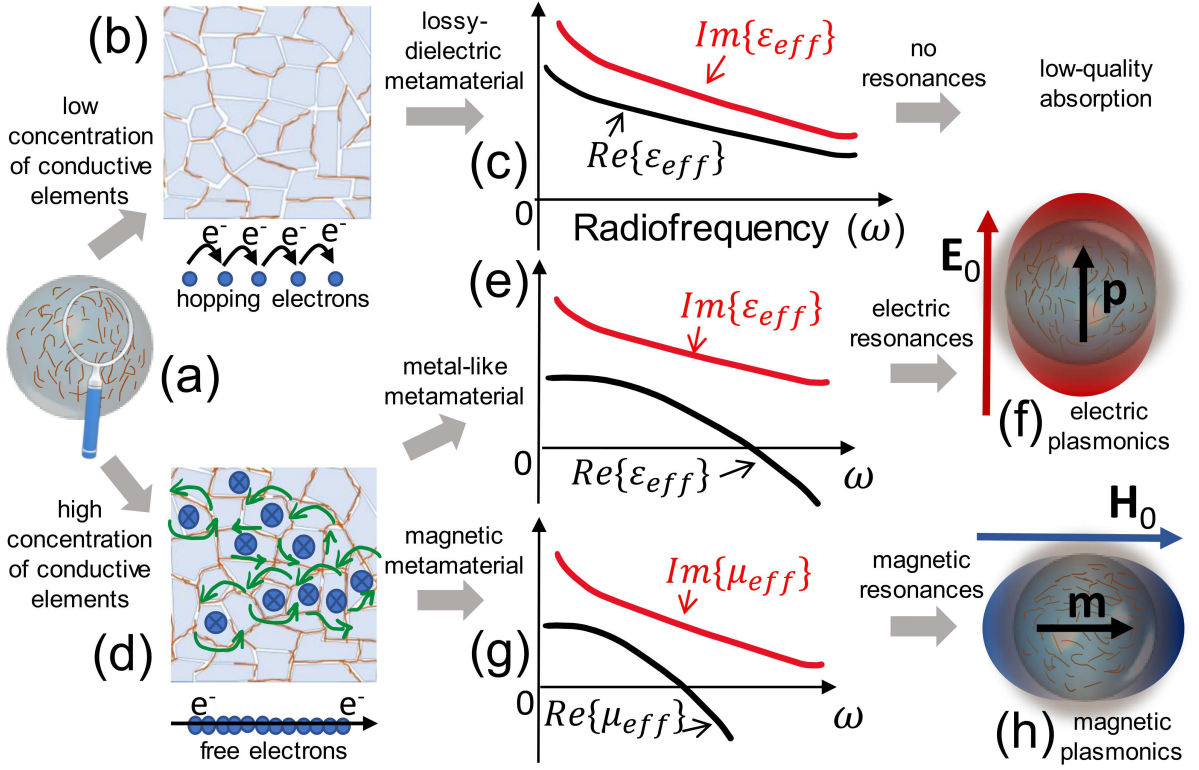"}
    \caption{ Radioplasmonics and mechanism to obtain electric and magnetic surface plasmons by using radiofrequency metamaterials. (a) The particle is made of a metamaterial composite represented as a single effective medium. In a simple case, a dielectric host medium (light blue) contains a concentration of conductive nanoelements (orange lines). Depending on this concentration, one can obtain (b) a disperse ``solution'' and hopping-electron conduction [inset in panel (b)], which results in (c) a lossy-dielectric metamaterial  with no resonances in absorption when the particle is illuminated, or (d) a concentrated solution near or beyond the percolation threshold, which forms a conductive network with metal-like conduction [free electrons are depicted by the inset in panel (d)]. (e) This metal-like behavior is represented by a negative-real part in the dielectric function for the effective metamaterial. When the particle is appropriately illuminated, an electric resonance can be excited. (f) An incident wave with electric field \(\mathbf{E}_0\) (red arrow) induces a dipole moment \(\mathbf{p}\) by means of the electric polarizability \(\alpha_E\). The resonance associated with \(\alpha_E\) is the dipolar electric surface plasmon, represented by a red spot in the scheme. (g) Negative real permeability can also appear as an attempt of the composite to oppose the incident magnetic fields, generating an inner magnetic response represented by the blue symbols and the green current loops in panel (d). Similarly to the electric phenomenon, a magnetic resonance can be excited with the appropriate illumination. (h) As a consequence, an incident wave with magnetic field \(\mathbf{H}_0\) (blue arrow) can induce a magnetic dipole moment \(\mathbf{m}\) via the  magnetic polarizability \(\alpha_M\). The resonance associated with the pole of \(\alpha_M\) is the dipolar magnetic surface plasmon, represented as a blue spot in the scheme. Moreover, the particle can be heated up by both electric and magnetic Joule heating when irradiated with radio waves. This effect is schematized as a pink cloud around the particle in panels (f) and (h).}
    \label{fig:fig1}
\end{figure*}

A MM particle consists of some conductive nanoelements immersed in a matrix or host medium [see Fig.~\ref{fig:fig1}(a)]. It constitutes a microcomposite, and,  for later purposes, the host medium has a low speed of sound. These nanoelements can be carbon nanotubes, graphene sheets, nanofibers, or even nanoparticle clusters; they typically define preferred directions for electrical conduction. When zoomed in, the particle shows that  three possibilities  may exist to function as an EM scatterer, depending on the concentration of the conductive elements in the matrix. A low concentration of elements [in orange, online version, Fig.~\ref{fig:fig1}(b)] in the matrix (in light blue) creates a composite below the so-called \textit{percolation threshold}, with poor electric conduction due to a hopping-electron mechanism [see inset below Fig. \ref{fig:fig1}(b)]. As a result, the entire MM behaves as a pure dielectric, with positive real and imaginary parts of the permittivity function  [see Fig.~\ref{fig:fig1}(c)]. This happens in any typical material in the RF domain. 

On the other hand, a high concentration of the conductor phase can produce a composite near or above the percolation threshold [Fig.~\ref{fig:fig1}(d)], with high electric conduction due to free-electron currents [see inset below Fig. \ref{fig:fig1}(b)]. As a consequence, a conduction net is established inside the MM, and one or two possibilities may arise: In the first case,  free-electron conduction creates a metal-like composite in the RF domain, and the real part of the permittivity becomes negative [Fig.~\ref{fig:fig1}(e)]. An oscillating electric field \(\mathbf{E}_0\) [see red arrow in Fig.~\ref{fig:fig1}(f)]  moves the electrons in the particle and induces an electric dipole moment \(\mathbf{p}\). In this way, an \textit{electric} surface plasmon (ESP) is excited at the appropriate wavelength [Fig.~\ref{fig:fig1}(f)], as  occurs in conventional plasmonics with  optical fields upon reaching the pole of the electric polarizability \(\alpha_E\)  (see Sec.  \ref{sub:EMR} for details). The red spots  correspond to the dipole electric field of the ESP [Fig.~\ref{fig:fig1}(f)]. 

In the second case, another effect can happen in addition to  electric conduction. A magnetic response of the MM may arise in the form of current loops around the net paths  (green arrows). This induced field in the MM responds to the magnetic field \(\mathbf{H}_0\) of the incident wave and couples with the electronic circulation (like in eddy currents); see the blue crosses in Fig.~\ref{fig:fig1}(d), which indicate a downward-induced field according to the current loops. In a form analogous to the negative real permittivity, negative real permeability can also occur as an attempt of the MM to oppose the disturbing external fields [see Fig.~\ref{fig:fig1}(g)]. A magnetic dipole moment appears as a response to the incident magnetic field [see blue arrow in Fig.~\ref{fig:fig1}(h)]. Next, a possible \textit{magnetic} surface plasmon (MSP) can be excited at the appropriate wavelength (see in Sec. \ref{sub:EMR} the condition to reach the pole in the magnetic polarizability \(\alpha_M\)). The blue spots in this case correspond to the dipole magnetic field of the MSP [Fig.~\ref{fig:fig1}(h)]. In addition, the particle can heat up by the electric and magnetic Joule effect. This phenomenon is schematized by a cloud spread out  around the particle [Figs. \ref{fig:fig1}(f) and \ref{fig:fig1}(h)].

As examples of realistic MMs, the simulated microparticles are composites made of three materials; an electric MM (EMM), a magnetic MM (MMM), and an acoustic or A-material. The A-material is a dielectric designed to admit the acoustic resonances in the MHz frequencies in combination with other materials. The speed of sound for A-material depends on the materials chosen for the composite. It can be as low as 60-100 m/s, as in rubber-like materials \cite{abraham-ekeroth_radioplasmonics_2021}, or reach orders of \(3-7 \times 10^3\) m/s like in this work (see more data in Sec.~\ref{sec:PhotoAcoustResults}).
\subsection{Electromagnetic response \label{sub:EMR}} 

We generalize a  procedure similar to that reported in Ref. \cite{abraham-ekeroth_radioplasmonics_2021} to include a magnetic response. A plane wave characterizes the spatial dependence of the EM radiation with electric and magnetic fields \(\mathbf{E}_0\) and \(\mathbf{H}_0\), respectively. Given that the minimum penetration depth is several times the particle size, the fields are treated as constant inside the particles and the quasistatic approximation is used,\cite{novotnyPrinciplesNanoOptics2006} which is justified if  \(R\ll\lambda\), where \(R\) is the particle radius and \(\lambda\) is the wavelength of the incident wave. The particle is then represented by its electric and magnetic dipole moments
\begin{eqnarray}
    \mathbf{p}&=&\epsilon_0\epsilon_{rb} \alpha_E \mathbf{E}_0,
\\
    \mathbf{m}&=&\alpha_M \mathbf{H}_0,
\end{eqnarray}
where \(\epsilon_{rb}\) is the relative dielectric function of the embedding medium. The polarizabilities are defined by \cite{albaladejoRadiativeCorrectionsPolarizability2010}
\begin{align}
    & \alpha_E = \left( \alpha_{E,0}^{-1}-\frac{ik_b^3}{6\pi} \right)^{-1},\quad \alpha_{E,0} = 3V\frac{\epsilon_{\text{r}}-\epsilon_{rb}}{\epsilon_{\text{r}}+2\epsilon_{rb}}, \label{eq:alphaE} \\
    & \alpha_M = \left( \alpha_{M,0}^{-1}-\frac{ik_b^3}{6\pi}  \right)^{-1},\quad \alpha_{M,0} = 3V\frac{\mu_{\text{r}}-\mu_{rb}}{\mu_{\text{r}}+2\mu_{rb}}, \label{eq:alphaM}
\end{align}
where \(k_b =\sqrt{\epsilon_{rb}\mu_{rb}} \omega'/c\), \(\omega'\) is the angular frequency, and \(c\) is the speed of light. \(k_b\) and \(\mu_{rb}\) are the wave number and relative permeability of the surrounding medium, respectively. \(V=4\pi R^3/3\) is the particle volume, and \(\epsilon_{\text{r}}\) and \(\mu_{\text{r}}\) are the relative dielectric and magnetic functions of the particle. Finally, \(\alpha_{E,0}\) and \(\alpha_{M,0}\) are  the quasi-static polarizabilities.

 If the composite is appropriately designed (i.e., the complex denominators of Eqs.~(\ref{eq:alphaE}) and (\ref{eq:alphaM}) are zero), the permittivities and permeabilities of the MMs used herein should at
 the desired spectral positions, fulfill one or both of the following Frölich conditions: (i) \(Re\{\epsilon_{\text{r}}(\omega')\}+2\epsilon_{rb}=0\), (ii) \(Re\{\mu_{\text{r}}(\omega')\}+2\mu_{rb}=0\). Generally, Frölich conditions are accurate approximations of the resonances' locations only for a relatively small or slowly varying imaginary part of the corresponding constitutive function \cite{maier_plasmonics_2007}. Nevertheless, Frölich's equations are not exact enough in the present work since the imaginary parts of the functions are either large or rapidly varying around the resonances. Instead, other conditions must be used \cite{fan_light_2014}. The dipolar electric and magnetic resonances can be located respectively near the minima of the functions (iii) \(\wp_E = (Re\{\epsilon_{\text{r}}(\omega')\}+2\epsilon_{rb})^2 + Im\{\epsilon_{\text{r}}(\omega')\}^2\) and (iv) \(\wp_M = (Re\{\mu_{\text{r}}(\omega')\}+2\mu_{rb})^2 + Im\{\mu_{\text{r}}(\omega')\}^2\); see the SM for more details. Satisfying one or both conditions of (iii)-(iv) being minima correspond to a single- or double-resonant behavior, one for each particle polarizability and, consequently, a resonant absorption in the corresponding spectrum. Each electric and magnetic resonance corresponds to the particle's excitation of the dipolar ESP and MSP.

Thus, the absorption cross section for the particle follows from the absorption by the electric and magnetic dipole moments:\cite{alcaraz_de_la_osa_new_2013}
\begin{equation}
    \label{eq:sigma_abs_tot}
    \sigma_{\text{abs}}=\sigma_{\text{abs,E}} + \sigma_{\text{abs,M}},
\end{equation}
where
\begin{align}
    & \sigma_{\text{abs,E}}=k_b\left(\text{Im}(\alpha_E)-\frac{k_b^3}{6\pi}|\alpha_E|^2\right), \label{eq:sigma_abs_E} \\ 
    & \sigma_{\text{abs,M}}=k_b\left(\text{Im}(\alpha_M)-\frac{k_b^3}{6\pi}|\alpha_M|^2\right). \label{eq:sigma_abs_M}
\end{align}
The field maps in this paper are calculated by a Mie code.\cite{craig_f_bohren_absorption_1998} Mie theory is also used to check the quasistatic approximation in the results.

From here on, all the results correspond to saline water as the surrounding medium with temperature \(T_0=36.6 \, ^\circ \text{C}\) and permittivity \(\epsilon_{rb}=73.3\).\cite{stogrynEquationsCalculatingDielectric1971} The composites' constitutive functions \(\epsilon_{\text{r}}(\omega')\) and \(\mu_{\text{r}}(\omega')\) are obtained from the Maxwell-Garnet model, using the data reported in \cite{chengRadioFrequencyNegative2017,li_preparation_2019} as input media in addition to the A-material. This procedure is detailed in the SM. Saline water with low salinity is used to mimic body tissues as a first approximation to the concepts under discussion below. \cite{abraham-ekeroth_radioplasmonics_2021,abraham-ekeroth_radioplasmonics_2021-4} The acoustic and mechanical properties also correspond to salty water. Please refer to the SM for more details.

\subsection{Photoacoustic phenomenon: Full solution \label{sub:PAPhFS}} 

The complete problem for the thermoacoustic generation of pressure is described by the following system of coupled equations in every medium involved,\cite{hatefAnalysisPhotoacousticResponse2015} assuming inviscid fluids:
\begin{align}
    \label{eq:TA_time}
    & \frac{\partial}{\partial t}\left(T-\frac{\gamma - 1}{\gamma \mathfrak{a}}p\right)= \frac{\bm\nabla \cdot (\kappa\bm \nabla T)}{\rho C_p} + \frac{Q}{\rho C_p}, \\ 
    & \left(\nabla^2 - \frac{\gamma}{v_{s}^2}\frac{\partial^2}{\partial t^2}\right)p = -\frac{\mathfrak{a} \gamma}{v_{s}^2}\frac{\partial^2T}{\partial t^2},
\end{align}
where \(T\) is the temperature, \(p\) is the pressure, \(\gamma\) is the specific heat ratio, \(\mathfrak{a}=\left({\partial p}/{\partial T}\right)_V\) is the pressure expansion coefficient at constant volume, \(\kappa\) is the thermal conductivity, \(\rho\) is the density, \(C_p\) is the heat capacity at constant pressure, \(v_\text{s}\) is the speed of sound, and \(Q\) is the energy deposited per unit time and unit volume in the particle, which transforms into heat. The inviscid-fluid approximation   is often used in photoacoustic imaging and may be close to what might be found in the laboratory.\cite{xu_photoacoustic_2006}
As a simplification, it is common to assume \(\gamma \approx 1\) for any typical water-like fluid. In any case, the condition \(
{\bm\nabla \cdot (\kappa\bm \nabla T)}/{(\rho C_p)}\ll{\partial T}/{\partial t}\) is a reasonable approximation for fast excitation in all media involved. Assuming also that  heating  occurs only in the particle's medium through the relation \(
{\partial T}/{\partial t} = {Q}/{(\rho C_p)}\), the PA generation by an arbitrary EM field is reduced to solve the following system of \textit{uncoupled} equations:
\begin{align}
    & \nabla^2 p_\text{s} - \frac{1}{v_{ss}^2}\frac{\partial^2 p_\text{s}}{\partial t^2}= -\frac{\beta} {C_p}\frac{\partial Q}{\partial t}, \label{eq:PA_time_in} \\ 
    & \nabla^2p_f - \frac{1}{v_{\text{sf}}^2}\frac{\partial^2 p_f}{\partial t^2} = 0, \label{eq:PA_time_out} 
\end{align}
where \(p_\text{s}\) (\(p_\text{f}\)) is the pressure inside (outside) the particle, \(\beta\) is the volumetric thermal expansion, and \(v_{\text{ss}}\) (\(v_{\text{sf}}\)) is the longitudinal speed of sound in the particle's (the embedding) medium. Here, transversal waves are not considered. The solution of the system~(\ref{eq:PA_time_in}) and (\ref{eq:PA_time_out}) is bound to the boundary conditions (i.e., continuity of the pressure and radial velocity for each medium), taking the spherical symmetry into consideration. In particular, as the particle is small enough, we assume that the heating source \(Q\) is homogeneous inside the particle and that the process is fast (i.e., \(\nabla^2 p_\text{s} \ll({1}/{v_{\text{ss}}^2}){\partial^2 p_\text{s}}/{\partial t^2}\). Then the inhomogeneous solution of Eq.~(\ref{eq:PA_time_in}) (i.e., \(P_0\)) is obtained from
\begin{equation}
        \frac{1}{v_{\text{ss}}^2}\frac{\partial^2 P_0}{\partial t^2}= \frac{\beta} {C_p}\frac{\partial Q}{\partial t}, \label{eq:source}
\end{equation}
such that \(p_\text{s} = p_{\text{sh}} + P_0\) is the complete solution of Eq.~(\ref{eq:PA_time_in}), where \(p_{\text{sh}}\) is the homogeneous part of the solution. \(p_\text{f}\) is the homogeneous solution of Eq.~(\ref{eq:PA_time_out}).

The complete solution for any EM excitation with source intensity \(I(t)\) can be built from the convolution product between a PA pressure generated by a monochromatic plane wave (MPW) and \(I(t)\):
\begin{align}
     p_{\text{s}}(r,t)& = p_{\text{s}}(r,t)|_{\text{MPW}} * \frac{I(t)}{I_0}   \nonumber \\ & =\int_{-\infty}^{\infty}p_{\text{s}}(r,t)|_{\text{MPW}}\frac{I(\hat{\tau})}{I_0}d\hat{\tau}, \label{eq:conv-time-ps-gral} \\
     p_{\text{f}}(r,t) &= p_{\text{f}}(r,t)|_{\text{MPW}} * \frac{I(t)}{I_0}  \nonumber \\ 
    & =\int_{-\infty}^{\infty}p_{\text{f}}(r,t)|_{\text{MPW}}\frac{I(\hat{\tau})}{I_0}d\hat{\tau}, \label{eq:conv-time-pf-gral} 
\end{align}
where \(I_0\), the initial intensity, is known. \(\hat{\tau}\) must be an adimensional retarded time to be coherent with the problem's physical dimensions. The solutions from  (\ref{eq:conv-time-ps-gral}-\ref{eq:conv-time-pf-gral}) are not immediate. However, this convolution product can be easily calculated using the Fourier transform. By taking the parity properties of the complex fields into account, the solution can be built from the frequency-domain functions as follows: 
\begin{align}
    & p_{\text{s}}(r,t) = \operatorname{Re}\left\{\mathscr{F}^{-1}\left[\hat{p}_{\text{s}}(r,\omega)\frac{\hat{I}(\omega)}{I_0}\right]\right\}, \label{eq:ps-conv-inv-FT} \\
    & p_{\text{f}}(r,t) = \operatorname{Re}\left\{\mathscr{F}^{-1}\left[\hat{p}_{\text{f}}(r,\omega)\frac{\hat{I}(\omega)}{I_0}\right]\right\}, \label{eq:pf-conv-inv-FT}
\end{align}
where the Fourier transform is defined as \(\mathscr{F}[f(t)] = \hat{f}(\omega)=\int_{-\infty}^{\infty}f(t)e^{i\omega t}dt\) such that \(\mathscr{F}^{-1}[\hat{f}(\omega)] = f(t)=\frac{1}{2\pi}\int_{-\infty}^{\infty}\hat{f}(\omega)e^{-i\omega t}d\omega\). 
 Specifically, the functions \(\hat{p}_{\text{s}}(r,\omega)\) and \(\hat{p}_{\text{f}}(r,\omega)\) are obtained from the frequency-domain solution to the PA generation under a MPW. To do this, the illumination takes the form \(I(t)=I_0e^{-i\omega t}\) such that \(Q(t)=\sigma_{\text{abs}}I(t)/V\). Using this \(Q(t)\) in Eq.~(\ref{eq:source}) gives \(P_0={i \beta v^2_{\text{ss}} \sigma_{\text{abs}}I_0}/{(\omega C_p V)}\). \(\hat{p}_{\text{s}}\) and \(\hat{p}_\text{f}\) are obtained from the corresponding Helmholtz equations derived from Eqs.~(\ref{eq:PA_time_in}) and (\ref{eq:PA_time_out}) under spherical symmetry. In general, they are expressed as a series in the spherical Bessel functions and the Legendre polynomials, but let us focus here only on the monopolar solutions, which are the ones that matter in the present study. The solution is constrained to the following boundary conditions: \cite{feuilladeAnderson1950Revisited1999}
 \begin{align}
        \label{eq:bc-freq-domain}
         \hat{p}_{\text{s}}|_{r=R} &= \hat{p}_{\text{f}}|_{r=R}, \\
         \frac{1}{\rho_\text{s}}\frac{\partial \hat{p}_\text{s}}{\partial r}\bigg|_{r=R} &= \frac{1}{\rho_\text{f}}\frac{\partial \hat{p}_\text{f}}{\partial r}\bigg|_{r=R}.
 \end{align}
Solving the system, we obtain \cite{yan_generation_2021}
 \begin{align}
         \hat{p}_\text{s}(r,\omega) &= P_0   \left[1 + \frac{\hat{v}\hat{\rho}h^{(1)}_1(k_fR)j_0(k_sr)}{h^{(1)}_0(k_fR)j_1(k_sR) - \hat{v}\hat{\rho}h^{(1)}_1(k_fR)j_0(k_sR) }\right],\nonumber\\ \label{eq:MPW-ps-freq-domain} \\ 
         \hat{p}_\text{f}(r,\omega) &= P_0 
     \frac{j_1(k_sR)h^{(1)}_0(k_fr)}{h^{(1)}_0(k_fR)j_1(k_sR) - \hat{v}\hat{\rho}h^{(1)}_1(k_fR)j_0(k_sR) }, \label{eq:MPW-pf-freq-domain}
\end{align}
where \(\hat{\rho}={\rho_\text{s}}/{\rho_\text{f}}\), and \(\hat{v}={v_{\text{ss}}}/{v_{\text{sf}}}\).  Equations~(\ref{eq:MPW-ps-freq-domain}) and  (\ref{eq:MPW-pf-freq-domain}) show that the acoustic monopole has resonance when the denominator of these expressions goes to zero. This is the acoustic resonance to be exploited in this work. Assuming illumination by a MPW, the time-domain solution is 
% %
 \begin{align}
    & p_{\text{s}}(r,t)|_{\text{MPW}} = -\frac{\Gamma \sigma_{\text{abs}}I_0}{\omega V} \label{eq:MPW-ps-time-domain}\\  
    &\times\text{Im}\Biggl\{ e^{-i\omega t}\left[ 1 + \frac{\hat{v}\hat{\rho}h^{(1)}_1(k_fR)j_0(k_sr)}{h^{(1)}_0(k_fR)j_1(k_sR) - \hat{v}\hat{\rho}h^{(1)}_1(k_fR)j_0(k_sR)} \right]\Biggr\},\nonumber  \\
    & p_{\text{f}}(r,t)|_{\text{MPW}} = -\frac{\Gamma \sigma_{\text{abs}}I_0}{\omega V}\label{eq:MPW-pf-time-domain} \\ 
    &\times\operatorname{Im}\Biggl\{e^{-i\omega t}\frac{j_1(k_sR)h^{(1)}_0(k_fr)}{h^{(1)}_0(k_fR)j_1(k_sR) - \hat{v}\hat{\rho}h^{(1)}_1(k_fR)j_0(k_sR)}\Biggr\}, \nonumber
 \end{align}
 so that the products in the time domain given in Eqs.~(\ref{eq:conv-time-ps-gral}) and (\ref{eq:conv-time-pf-gral}) are now defined. Note that the formulation is written in terms of \(\Gamma={\beta v^2_{\text{ss}}}/{C_p}\), which is the well-known Grüneisen coefficient.\cite{zhou_tutorial_2016}

Observe that Eqs.~(\ref{eq:ps-conv-inv-FT}) and (\ref{eq:pf-conv-inv-FT}) are general. For example, in the limit of an extremely fast pulse, \(I(t)=I_0\tau\delta(t-t_0)\), where \(\tau\) is the characteristic width of such pulse and \(t_0\) is the initial time from the pulse's maximum, then \(\hat{I}(\omega)/I_0=\tau\exp{i\omega t_0}\), and \(p_{\text{s}}(r,t)|_{\text{MPW}}\) and  \(p_{\text{f}}(r,t)|_{\text{MPW}}\) are recovered by the definition of the inverse Fourier transform of a MPW.

 At this point, a Gaussian pulse with full width at half maximum \(\tau\) becomes the EM source \(g(t)=\frac{I(t)}{\tau I_0}\) corresponding to the normal distribution 
 \begin{equation}
        \label{eq:gaussian_time}
        g(t)=\frac{2\sqrt{\ln(2)}}{\sqrt{\pi}\tau}\exp\left[{-4\frac{\ln(2)(t-t_0)^2}{\tau^2}}\right],
 \end{equation}
 where \(t_0\) is defined above. If the Fourier transform of the Eq.~(\ref{eq:gaussian_time}) is considered, a full time-domain solution for the PA pressure generated is obtained from Eqs.~(\ref{eq:ps-conv-inv-FT}-\ref{eq:pf-conv-inv-FT}), letting:
 \begin{align}
         p_{\text{s}}(r,t) &= \frac{\tau}{\pi}\operatorname{Re}\left\{\int_{0}^{\infty}\hat{p}_{\text{s}}(r,\omega)e^{-\frac{\tau^2\omega^2}{16\ln(2)}}e^{-i\omega t}d\omega\right\}, \label{eq:ps-conv-freq-pulse} \\
         p_{\text{f}}(r,t) &= \frac{\tau}{\pi}\operatorname{Re}\left\{\int_{0}^{\infty}\hat{p}_{\text{f}}(r,\omega)e^{-\frac{\tau^2\omega^2}{16\ln(2)}}e^{-i\omega t}d\omega\right\}.
     \label{eq:pf-conv-freq-pulse} 
 \end{align}
Note that, in the limit \(\tau \rightarrow 0\), \(I(t)=I_0\tau\delta(t-t_0)\), solutions~(\ref{eq:ps-conv-freq-pulse}) and (\ref{eq:pf-conv-freq-pulse}) go to zero because the pulse has  finite energy [\(g(t)\) is a normalized function]. 

Explicitly, inserting Eqs.~(\ref{eq:MPW-ps-freq-domain}) and (\ref{eq:MPW-pf-freq-domain}) into Eqs.~(\ref{eq:ps-conv-freq-pulse}) and (\ref{eq:pf-conv-freq-pulse}), the PA pressure in the time-domain for a Gaussian pulse can be rewritten as
\begin{align}
    & p_\text{s}(r,t) = -\frac{\sigma_{\text{abs}}(\omega')\Gamma F}{\pi V} \operatorname{Im}\Biggl\{\int_{0}^{\infty}e^{-iq\hat{\tau}}e^{-\frac{\tau^2q^2v^2_{\text{ss}}}{16R^2\ln(2)}} \Biggr. \nonumber \\ 
    & \times\Biggl. \left[\frac{1}{q} + \frac{R}{r}\frac{\hat{\rho}\sin(\frac{qr}{R})/q^2 - i\hat{v}\hat{\rho}\sin(\frac{qr}{R})/q}{\left[(1-\hat{\rho})\frac{\sin(q)}{q} - \cos(q) + i\hat{v}\hat{\rho}\sin(q)\right]} \right]dq \Biggr\}, \label{eq:ps_time_sol} \\
     p_\text{f}(r,t)& = -\frac{\sigma_{\text{abs}}(\omega')\Gamma F}{\pi V}\frac{R}{r}  \nonumber \\  
    & \times \operatorname{Im}\left\{\int_{0}^{\infty} e^{-iq\hat{\tau}}e^{-\frac{\tau^2q^2v^2_{\text{ss}}}{16R^2\ln(2)}} P(q) dq  \right\} \label{eq:pf_time_sol}, \\
     P(q)& = \frac{\left[\sin(q)-qcos(q)\right]/q^2}{\left[\left(1-\hat{\rho}\right)\frac{\sin(q)}{q} - \cos(q) + i\hat{v}\hat{\rho}\sin(q)\right]}, \label{eq:argument_pf_time_sol}
\end{align}
where \(r \neq 0\), and the acoustic variable \(q=\omega R/v_{\text{ss}}\) is defined as per Ref.~\onlinecite{wang_photoacoustic_2017}, which involves the acoustic angular frequency \(\omega\). In this complete version of the PA response under Gaussian illumination, note that \(\sigma_{\text{abs}}(\omega')\) depends in general  on the driving frequency \(\omega'\) (in general \(\neq \omega\)) of the EM field. \(F=I_0\tau\) is the pulse fluence, \(r\) is the radial distance from the particle's center, and \(\hat{\tau}={v_{\text{ss}}}[t-t_0 - {(r-R)}/{v_{\text{sf}}}]/{R}\). 

From this complete solution, it is clear that at least two kinds of resonances can couple; one is any \textit{implicit} EM resonance in \(\sigma_{\text{abs}}\), the other is the acoustic resonance corresponding to the excitation of any monopolar mode. The concept of mixing different resonance types in a novel PA phenomenon is the core of the present study. 
\section{Numerical Results and discussion} 

\subsection{Excitation of radioplasmons} 

The discussion that follows concerns a \(10\ \mu\)m particle made of three types of realistic materials. The first material is a magnetic-electric composite: a Y\(_{3}\)Fe\(_{5}\)O\(_{12}\) matrix (yttrium iron garnet, or YIG)  filled with Ag nanoparticles. This is the MMM that endows the particle with magnetic properties.\cite{li_preparation_2019} The second compound is the EMM, namely, an Al\(_{2}\)O\(_{3}\) matrix containing multiwalled carbon nanotubes (MWCNTs) as inclusions\cite{chengRadioFrequencyNegative2017} to give the electric properties. The final material is a high dielectric that has a low speed of sound; it serves as an A material (or simply material ``A'') for later purposes. Its dielectric constant is \(\epsilon_{\text{rA}}=250\) with a negligible imaginary part, and \(\mu_{\text{rA}}=1\). As a realistic example, this material can be a polyaniline-based polymer.\cite{huang_high-dielectric-constant_2003,shi_dielectric_2018} For more details of the data used in the simulations, see the  SM. 

Although nowadays very likely to be manufacturable, we do not discuss how to fabricate microscale composites \cite{chen_three-dimensional_2011,kaufman_structured_2012,yin_electromagnetic_2014,chen_negative_2016,javadi_stretching_2017,sun_overview_2018,messina_graphene_2018,teng_comparison_2018,sunTunableNegativePermittivity2019,messina_simultaneous_2020} but rather show the consequences of assuming microspheres.
\begin{figure}
    \centering
    \includegraphics[width=\columnwidth]{"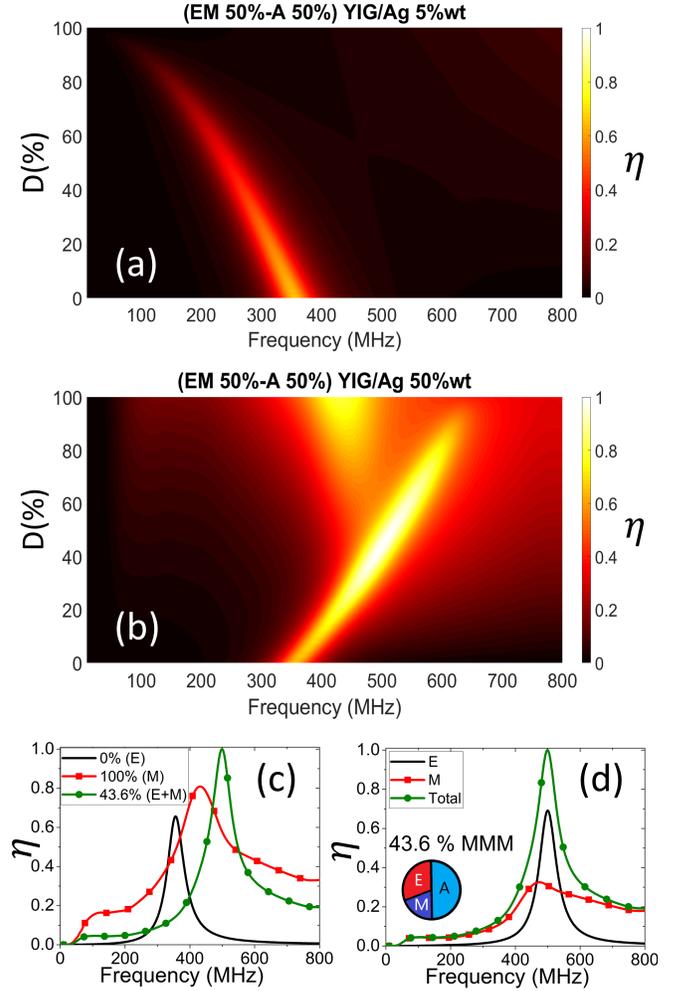"}
    \caption{ Normalized electromagnetic absorption \(\eta\) in composites based on [Ag(\(X\))/YIG]\textsubscript{D/2}[MWCNT(11\%)/Al\(_2\)O\(_3\)]\textsubscript{50\%-\(D\)/2}[A]\textsubscript{50\%}. (a), (b) Maps as a function of the filling fraction \(D\) in percent of the magnetic metamaterial and the frequency. (a) For Ag at \(X=5\%\), under the percolation threshold, \(\mu_{\text{r}}(\text{MWCNT}(11\%)/\text{Al}_2\text{O}_3)=1\). (b) For Ag at \(X=50\%\), beyond the percolation threshold, \(\mu_{\text{r}}(\text{MWCNT}(11\%)/\text{Al}_2\text{O}_3)=0.8\). The maximum absorption is found for \(D=43.6\%\) and at a frequency of 499.5 MHz. (c) The absorption of this composite is compared with that of pure MWCNT(11\%)/Al\(_2\)O\(_3\) and pure Ag(50\%)/YIG. (d) Electric and magnetic contributions for \(D=43.6\%\), showing the addition of the electric and magnetic surface plasmons. The particle composition is schematized with a pie chart, where E (M) stands for EMM (MMM) for simplicity. This is also used in the legends of panels (c) and (d).}
    \label{fig:fig2}
\end{figure}
As a starting point, the EM absorption by the particle is studied as a function of the particle composition and driving frequency \(
{\omega'}/{2\pi}\). The normalized absorption \(\eta\) is the absorption cross section \(\sigma_{\text{abs}}\) scaled by the factor \(\sigma_{\text{abs,max}}=0.42 \ \mu \text{m}^2\) (see Fig.~\ref{fig:fig2}). For convenience, half of the composite corresponds to the A-material, which acts as a host or matrix. The remaining 50\% of the composite is left for the EM MMs whose properties are varied. The MMM can be written as [Ag(\(X\))/YIG]\(\textsubscript{D/2}\), where \(D\) is its concentration in percent and acts as the guest that fills  the host [MWCNT(11\%)/Al\(_2\)O\(_3\)]\textsubscript{50\%-\(D\)/2}. Here, \(X\) is the concentration in percent of Ag fillers in YIG. As reported in Ref.~\cite{li_preparation_2019}, \(X=5\%\) is below the percolation limit in Ag/YIG but shows positive permeability (\(\mu_{\text{r}} \approx 5\)) for this MM [see Fig.~\ref{fig:fig2}(a)]. Note how an ESP appears around \(355\) MHz for \(D=0\%\) due to the EMM. As \(D\) increases, the ESP ``redshifts'' and vanishes since this MMM does not support EM resonances. In Fig.~\ref{fig:fig2}(a), the EMM is calculated with \(\mu_{\text{r}}=1\). 

Very different behavior is found for \(X=50\%\) in the magnetic composite, which is beyond the percolation limit for Ag/YIG\cite{li_preparation_2019} [see Fig.~\ref{fig:fig2}(b)]. In this case, the ESP due to the presence of EMM is blueshifted. In addition, the MMM satisfies the condition of a MSP; it appears as a spot with a maximum  around \(432\) MHz for \(D=100\%\). More importantly, however, the maximum absorption is found for \(D=43.6\%\), 499.5 MHz because both the ESP and MSP overlap when the three materials coexist. For Fig.~\ref{fig:fig2}(b), the EMM is calculated with \(\mu_{\text{r}}=0.8\) since this composite can have some magnetic activity. This maximum value corresponds to \(\sigma_{\text{abs,max}}=0.42 \ \mu \text{m}^2\) so that \(\eta=1\).

By comparing the absorption for the optimal condition \(D=43.6\%\) with that for pure EMM and MMM [Fig.~\ref{fig:fig2}(c)], the particle resonances can be tuned by their composition. The particle can be electric, magnetic, or electric-magnetic as required for different purposes. Note that Ag(50\%)/YIG is already an electric-magnetic material; observe that the red line with square symbols also has  a low-energy peak around \(100\) MHz. However, this peak is so weak and out of resonance compared with the MSP that another EMM is needed to strengthen the whole composite at the desired spectral location. To appreciate the overlapping effect of both the ESP and MSP at the optimal condition \(D=43.6\%\), the individual absorptions due to the electric and magnetic dipoles are calculated by using Eqs.~(\ref{eq:sigma_abs_E}) and (\ref{eq:sigma_abs_M}) [see the black solid curve and the red  curve of squares in Fig.~\ref{fig:fig2}(d), respectively]. The sum of the curves gives the total absorption for the particle, shown as a green curve of circles. Note how both types of plasmons are perfectly overlapped to enhance the absorption efficiency at \(499.5\) MHz. This kind of superposition can be designed for any MM particle at any frequency if the parameters of the mixture are carefully chosen. 
\begin{figure}
    \centering
    \includegraphics[width=1\linewidth]{"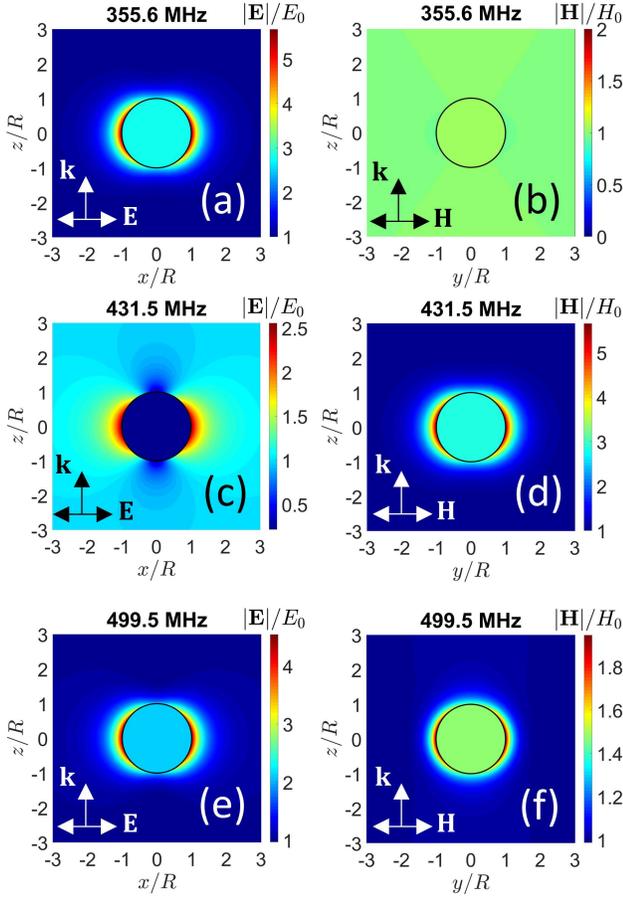"}
    \caption{ Near-field maps around a 10~\(\mu\)m particle showing plasmonic resonances under plane-wave illumination. The particle composite is [Ag(\(X\))/YIG]\textsubscript{\(D\)/2}[MWCNT(11\%)/Al\(_2\)O\(_3\)]\textsubscript{50\%-\(D\)/2}[A]\textsubscript{50\%}: (a), (c), (e) electric field, (b), (d), (f) magnetic field. The arrows indicate the wave vector \(\mathbf{k}\) and field directions. (a), (b) Electric surface plasmon for \(D=0\)\% or filling fraction = 0. (c), (d) Magnetic surface plasmon for  \(D=100\)\% or filling fraction = 1. (e), (f) Electric and magnetic surface plasmons for \(D=43.6\)\% or filling fraction = 0.436. The frequencies calculated appear above each panel.}
    \label{fig:fig3}
\end{figure}
In the remainder of this work, we focus on [Ag(50\%)/YIG] as the MMM. 

We now analyze the nature of these resonances by calculating near-field maps at the spectral locations of the SPs (Fig.~\ref{fig:fig3}). The maps show the magnitude of each field scaled to the respective values of the incident wave; see on top of each color bar. The left (right) column corresponds to the electric (magnetic) field. The incident wave approaches from the bottom, as the inset arrows indicate for the incident wave vector. Figures~\ref{fig:fig3}(a) and \ref{fig:fig3}(b) show the field patterns for \(D=0\%\) or pure EMM; the ESP is excited, showing the typical distribution of a SP for the particle [Fig. \ref{fig:fig3}(a)], and there is no appreciable magnetic activity [Fig. \ref{fig:fig3}(b)]. Conversely, for a pure MMM, namely, \(D=100\%\), there are both electric and magnetic responses [Figs.~\ref{fig:fig3}(c) and \ref{fig:fig3}(d)]. This phenomenon occurs because, as mentioned above, Ag(50\%)/YIG is  an electric-magnetic MM. Although the electric response is not resonant at \(431.5\) MHz, the magnetic response is resonant. Figures~\ref{fig:fig3}(e) and \ref{fig:fig3}(f) show the results for the optimal condition \((D=43.6\%\), 499.5 MHz). In this case, both types of patterns clearly show dipolar resonances corresponding to the ESP and MSP. Note that, in all the maps, the fields inside the particle are constant and the fields outside the particle have dipolar patterns: this confirms that the particle forms a dipolar scatterer given its small size compared with the underwater EM wavelength in the RF domain. The reader can find more details about the dipole resonances and penetration of RF waves in the studied particles in the SM. In addition, the skin depths in human tissues around 100-1000 MHz are compared to evaluate the biomedical prospects of the proposed method, see the SM.

\subsection{Excitation of acoustic resonances: Double- and triple-resonance frameworks and optimized photoacoustic pressure \label{sec:PhotoAcoustResults}} 

In what follows, we focus on the outer pressures, namely, waves that propagate in the surrounding medium. A calculation of the pressure at the particle's center is shown in the SM for the best result reported here. 

To better visualize the functions, all  PA calculations are done with the parameter \(t_0=3\tau\). The functions are normalized for simplicity. To conform to safety considerations for human exposure to RFs, the intensity in air was held constant at 4 GW/m\(^2\), which translates into an electric field in air of \(1.736\times 10^6\) V/m for a few nanoseconds.\cite{IEEEStandardSafety2019,klauenberg_sto_2018}

\begin{figure}%[h!]
    \centering    \includegraphics[width=1\linewidth]{"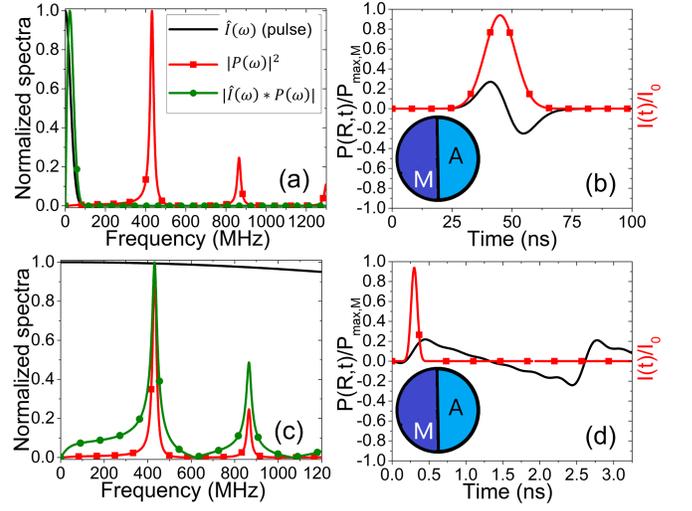"}
    \caption{ Photoacoustic generation of pressure waves using a \(10~\mu\)m particle made of [Ag(50\%)/YIG]\textsubscript{50\%}[A]\textsubscript{50\%} under pulse illumination and magnetic plasmon excitation. The functions are normalized for simplicity: (a), (b) for a ``long'' pulse of \(\tau=15\) ns; (c), (d) for ``short'' pulse of \(\tau=0.1\) ns. (a), (c) Comparison of the Fourier spectra of the pulse, the acoustic monopole intensity in water, and the product of the pulse and the monopole pressure. (b), (d) Pressure in the time domain at the particle's surface (\(r=R\)), scaled to  \(P_{\text{max,M}}=1016.6\) Pa for later purposes. The particle composition is schematized in the pie chart.}
    \label{fig:fig4}
\end{figure}

\begin{figure}
    \centering
    \includegraphics[width=1\linewidth]{"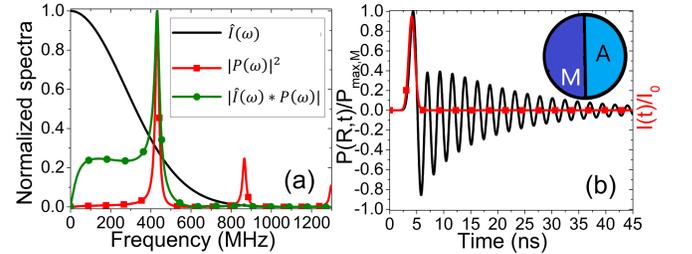"}
    \caption{(Color online) Idem Fig.~\ref{fig:fig4} under pulse illumination constrained to the condition imposed by Eq.~\ref{eq:pulse-acoust-mode-matching} (\(\tau \approx 1.36\) ns) where \(f_\text{A}=431.5\) MHz is the frequency of the acoustic resonance that matches the magnetic surface plasmon; magnetic \textit{double-resonance} phenomenon.}
    \label{fig:fig5}
\end{figure}

First, a novel concept of magnetic PA generation is illustrated. The results for a pure magnetic particle made of [Ag(50\%)/YIG]\textsubscript{50\%}[A]\textsubscript{50\%} are shown in Figs.~\ref{fig:fig4} and \ref{fig:fig5}. Moreover, a double-resonance phenomenon is explored in Fig.~\ref{fig:fig5}, namely, the simultaneous excitation of both the MSP and the acoustic monopole to enhance the magnetic PA generation. In contrast, Fig.~\ref{fig:fig4} shows the usual situations where the SP is excited but does not match the acoustic resonance. The particle parameters for the calculations performed in Figs.~\ref{fig:fig4} and \ref{fig:fig5} are \(C_{\text{p,eff}}=430.3\) J\,K\(^{-1}\)\,kg\(^{-1}\), \(\rho_{\text{eff}}=4422.5\) kg/m\(^3\), \(\nu_{\text{eff}} = 0.483\), \(E_{\text{eff}} = 7.93\) GPa, \(\alpha_{\text{eff}} = 1.064\times 10^{-4}\)~K\(^{-1}\). To obtain these parameters for the particle, the A-material must have the following thermomechanical properties: \(C_{\text{p,A}}=1000\) J\,K\(^{-1}\)\,kg\(^{-1}\), \(\rho_{\text{A}}=1100\) kg/m\(^3\), \(\nu_{\text{A}} = 0.475\), \(E_{\text{A}} = 3.124\) GPa, \(\alpha_{\text{A}} = 3.2\times 10^{-4}\) K\(^{-1}\). For more details, please refer to the data in the SM. 

Figure~\ref{fig:fig4} shows two situations:\ one calculated using a relatively long pulse of width \(\tau=15\) ns with a fluence \(F=60\) J/m\(^2\) [Figs. \ref{fig:fig4}(a) and \ref{fig:fig4}(b)], and another using a relatively short pulse of \(\tau=0.1\) ns and \(F=0.4\) J/m\(^2\) [Figs. \ref{fig:fig4}(c) and \ref{fig:fig4}(d)]. Figures~\ref{fig:fig4}(a) and \ref{fig:fig4}(c) show frequency spectra whereas Figs.~\ref{fig:fig4}(b) and \ref{fig:fig4}(d) show time-domain signals.

The coupling between the electromagnetic and the acoustic resonances can be visualized through Figs.~\ref{fig:fig4}(a) and \ref{fig:fig4}(c).  Figure~\ref{fig:fig4}(a) shows  how far off resonance are the excitation pulse and the acoustic oscillations since the monopole mode has a much higher frequency than the pulse width in the frequency domain (compare the black solid curve with the red curve of squares). The acoustic resonance is shown by the nondimensional magnitude \(|P|^2 = {|\hat{p}_\text{f}(R,\omega)|^2}/{P^2_0}\), which is the relative intensity of the acoustic mode. The fact that \(|P|^2\) scales inversely with \(P^2_0\) makes the result independent of the PA source of excitation. The value of \(\tau=2\sqrt{2 \ln 2}\sigma\), with \(\sigma\) being the standard deviation of the pulse (a normal distribution), has an associated width in the frequency domain of \(\sigma_{\omega}=1/\sigma\) that does not  include the particle's acoustic resonance in the integration [see Fig.~\ref{fig:fig4}(a)]. As a consequence, the product \(|\hat{I}\cdot P|\) between the pulse and the acoustic resonance is relatively weak, see the green curve of circles in Fig. \ref{fig:fig4}(a). As a result, the time-domain pressure is relatively low and damped [see the black solid curve in Fig.~\ref{fig:fig4}(b)], resembling the typical shape of the PA pressure generated in the optical range with the usual plasmonics. The result is independent of the type of SP being excited.

To compare the relative phases between the incoming pulse and the PA pressure, the red curve with symbols shows the EM pulse following the intensity scale on the right. This plot shows the lack of phase matching between the signals.

Figures~\ref{fig:fig4}(c) and \ref{fig:fig4}(d) present a different situation. The shorter the pulse in time [i.e., \(I(t)\rightarrow I_0\delta(t)\)], the wider the pulse \(\hat{I}(\omega)\) in the frequency domain. In Fig.~\ref{fig:fig4}(c), \(\hat{I}(\omega)\) is very wide and ``contains'' several monopolar modes, so they are excited as expected. However, as the pulse is normalized, a finite amount of energy is distributed over all the modes. As a result, the coupling with the first monopolar mode is weak, and the product \(|\hat{I} \cdot P|\) is relatively low and spread in all the first modes, see the green curve with circles in Fig.~\ref{fig:fig4}(c). Figure~\ref{fig:fig4}(d) shows again an inadequate phase matching between the signals and the superposition of higher-order monopoles in the pressure (small ripples). As mentioned in Sec. \ref{sub:PAPhFS}, the short pulses are not useful because the pressures weakens as \(\tau \rightarrow 0\). 

Conversely, Fig. \ref{fig:fig5} shows a perfect coupling between the resonances. The condition for the best phase matching between the pulse and the acoustic mode is that the pulse width in the time domain should equal the half period of the acoustic mode. In other words,
\begin{align}
     \Delta t_{\text{pulse}} &= \frac{T_\text{A}}{2} \nonumber \\ 
    & \approx 2\sigma = \frac{1}{2f_\text{A}} ,\nonumber \\ 
     \implies \tau &= \frac{\sqrt{2\ln 2}}{2f_\text{A}}. \label{eq:pulse-acoust-mode-matching}
\end{align}
In this expression, \(\Delta t_{\text{pulse}}\) is the total pulse duration, which is approximately  \(2\sigma\), \(T_\text{A}\) is the period of the acoustic monopole, and \(f_\text{A}\) is its frequency. As in Fig.~\ref{fig:fig4}, the acoustic monopole is ``prepared'' to resonate at the same spectral location as the MSP. Using the "resonance pulse matching condition" imposed by Eq.~(\ref{eq:pulse-acoust-mode-matching}) gives \(\tau=1.36\) ns, so the fluence is \(F=5.46\) J/m\(^2\). In this case, the pulse width in the frequency domain is large enough to include only the location of the first acoustic monopole [Fig.~\ref{fig:fig5}(a)]. The maximum possible coupling of the pulse occurs with the strongest monopole; see the relatively high values of \(|\hat{I} \cdot P|\) at low frequencies (green curve of circles). Note also from Fig.~\ref{fig:fig5}(a) that another monopolar resonance is  excited (see the red curve of squares around 850 MHz). However, this secondary resonance is poorly included in the convolution and does not count for the effect. As a result of  Eq.~(\ref{eq:pulse-acoust-mode-matching}), a resonant signal for the PA pressure appears in the time domain [Fig.~\ref{fig:fig5}(b)]. Figure~\ref{fig:fig5}(b) also shows the perfect phase matching between the pulse and the acoustic mode to  excite the mode with maximum efficiency. The signal is enhanced and reaches the top of the scale, namely, \(P_{\text{max,M}}=1016.6\) Pa, being on the order of PA generation due to gold nanoparticles using optical (electric) plasmonics, if one takes into consideration that the pulse fluence in the current case is in the order of a few J/\(m^2\) .\cite{prostPhotoacousticGenerationGold2015,kumarSimulationStudiesPhotoacoustic2018,kumar_pulsed_2019} The typical values of optical fluence varies between 10-100  J/\(m^2\), see for instance Refs.~ \cite{kumar_pulsed_2019,hatefAnalysisPhotoacousticResponse2015}. The maximum pressure (in absolute value) is in Fig.~\ref{fig:fig5}(b) \(\sim 3.7\) times higher than that for Fig.~\ref{fig:fig4}b and \(\sim 4.2\) times higher than that for Fig.~\ref{fig:fig4}d.

\begin{figure}
    \centering
    \includegraphics[width=0.99\linewidth]{"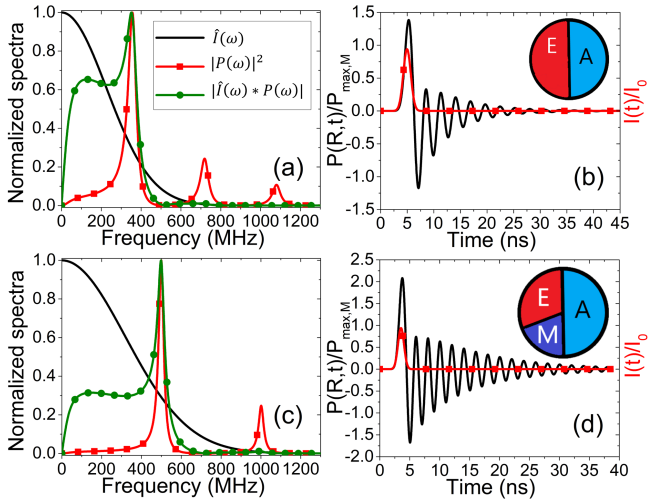"}
    \caption{(a), (b)\ Electric \textit{double-resonance} vs (c), (d) \textit{triple-resonance} phenomenon. In panels (a) and (b), the particle is made of [MWCNT(11\%)/Al\(_2\)O\(_3\)]\textsubscript{50\%}[A]\textsubscript{50\%} at the electric plasmon resonance. The incoming pulse obeys the resonance pulse matching condition,  given by Eq.~(\ref{eq:pulse-acoust-mode-matching}), i.e. (\(\tau \approx 1.66\) ns), where \(f_\text{A}=355.6\) MHz is the resonance acoustic frequency that matches the electric surface plasmon. In panels (c) and (d), the particle is made of [Ag(50\%)/YIG]\textsubscript{21.8\%}[MWCNT(11\%)/Al\(_2\)O\(_3\)]\textsubscript{28.2\%}[A]\textsubscript{50\%}. Now Eq.~(\ref{eq:pulse-acoust-mode-matching}) gives \(\tau \approx 1.18\) ns, and \(f_\text{A}=499.5\) MHz matches the resonance of the electric and magnetic surface plasmon.}
    \label{fig:fig6}
\end{figure}

Naturally, a comparison of the above case with a ``pure electric'' case is interesting because the latter  corresponds to  typical photoacoustics. Following the example of this work, a particle made of [MWCNT(11\%)/Al\(_2\)O\(_3\)]\textsubscript{50\%}[A]\textsubscript{50\%} is simulated with the conditions to achieve the double-resonance phenomenon by matching the ESP with the acoustic monopole [see Figs.~\ref{fig:fig6}(a) and \ref{fig:fig6}(b]. Thus, the spectral location of the first acoustic mode is set at 355.6 MHz, the same location as the ESP [see red curve of squares in Fig.~\ref{fig:fig6}(a)]. The particle parameters for the calculations performed in Figs.~\ref{fig:fig6}(a) and \ref{fig:fig6}(b) are \(C_{\text{p,eff}}=786.77\) J\,K\(^{-1}\)\,kg\(^{-1}\), \(\rho_{\text{eff}}=1934\) kg/m\(^3\), \(\nu_{\text{eff}} = 0.476\), \(E_{\text{eff}} = 3.397\) GPa, \(\alpha_{\text{eff}} = 1.56\times 10^{-4}\) K\(^{-1}\). To achieve these parameters for the particle, the acoustic material is assumed to have the same thermomechanical properties as those used for Figs.~\ref{fig:fig4} and \ref{fig:fig5}, except for \(\nu_{\text{A}} = 0.4837\) and \(E_{\text{A}} = 1.543\) GPa.

Three acoustic monopoles of decreasing intensities for increasing energies exist in the range plotted. However, the problem is concerned with the first monopole. Following Eq.~(\ref{eq:pulse-acoust-mode-matching}), \(\tau \approx 1.66\) ns is obtained, meaning a fluence of \(F=6.62\) J/m\(^2\). Thus, \(|\hat{I} \cdot P|\) is high at low frequencies, as shown by the green curve of circles in Fig.~\ref{fig:fig6}(a). Consequently, the PA signal in the time domain is resonant and enhanced compared with the case for ``magnetic double-resonance.'' Given that the fluence is similar to that for the ``pure magnetic'' case but a little higher, both results are comparable.

Finally, a triple-resonance phenomenon is illustrated in Figs.~\ref{fig:fig6}(c) and \ref{fig:fig6}(d). The composite [Ag(\(X\))/YIG]\textsubscript{\(D\)/2}[MWCNT(11\%)/Al\(_2\)O\(_3\)]\textsubscript{50\%-\(D\)/2}[A]\textsubscript{50\%} is assumed for the particle with \(D=43.6\%\), as studied above. Now the particle is designed to have the first acoustic monopole at \(499.5\) MHz, the same spectral location as both the MSP and ESP of the composite [Fig.~\ref{fig:fig6}(c)]. The particle parameters for the calculations performed in Figs.~\ref{fig:fig6}(c) and \ref{fig:fig6}(d) are \(C_{\text{p,eff}}=559.15\) J\,K\(^{-1}\)\,kg\(^{-1}\), \(\rho_{\text{eff}}= 3018.71\) kg/m\(^3\), \(\nu_{\text{eff}} = 0.475\), \(E_{\text{eff}} = 10.588\) GPa, \(\alpha_{\text{eff}} = 1.558\times 10^{-4}\) K\(^{-1}\). To obtain these parameters for the particle, the acoustic material is assumed to have the same thermomechanical properties as before, except for \(\nu_{\text{A}} = 0.4842\) and \(E_{\text{A}} = 4.93\) GPa.

This time, \(|\hat{I} \cdot P|\) is maximal for \(\tau \approx 1.18\) ns, in agreement with Eq. (\ref{eq:pulse-acoust-mode-matching}) [see green curve of circles in Fig.~\ref{fig:fig6}(c)]. This value of \(\tau\) means a pulse fluence of \(F=4.71\) J/m\(^2\). The function \(|\hat{I} \cdot P|\) is lower in intensity at low energies than that function for Fig.~\ref{fig:fig6}(a) (the scaling factor is practically equal in both cases). However, the PA pressure in the time domain for the triple-resonant particle is enhanced [Fig.~\ref{fig:fig6}(d)], even though the pulse fluence is \(71 \%\) of \(6.62\) J/m\(^2\). The maximum for this signal is more than twice the maximum for the ``pure magnetic'' particle. In addition, the signal lasts longer than the signal for the pure magnetic particle. Comparing the resonant values of \(\tau\) and given an almost constant fluence around \(5\) J/m\(^2\), the triple-resonance signal at the particle surface is about 1.1 times greater than the magnetic double-resonance signal and 35.6 times greater than the electric double-resonance signal at \(t \approx 29\tau\). The resonant signal for the electric particle is shorter  because the energy of the acoustic mode is the least  of all cases, namely \(f_\text{A}=355.6\) MHz. Conversely, the lower the frequency, the higher the body penetration. However, a longer-lasting pressure  supposes an advantage to detect the signal with an ultrasound transducer, except for the possible attenuations due to ultrasound absorption in the outer medium.\cite{huangBloodVesselImaging2018}

Now we compare the results coming from conventional plasmonics with the present results. This comparison is rough and should be taken with caution: in real applications, more parameters must be considered, like the number of particles per unit volume, for instance \cite{prostPhotoacousticGenerationGold2015}. In Ref.~\cite{prostPhotoacousticGenerationGold2015}, a PA signal generated by a 40 nm gold nanoparticle under plasmon excitation at \(532\) nm was calculated with \(\tau\) in the ns range (5 ns), using a fluence equivalent to \(F=1\) J/m\(^2\). The absorption cross-section reported was \(\sigma_{\text{abs}}=3.3 \times 10^{-15}\) m\(^2\). The pressure signal was calculated at 1 mm from the particle's center, giving a maximum of 1.25 mPa. In the example given here due to triple resonance, the pressure 1 mm away from the particle when \(F=1\) J/m\(^2\) reaches \(2.25\) Pa (i.e., 1800 times the maximum achieved in Ref. \cite{prostPhotoacousticGenerationGold2015} with conventional plasmonics). It is also known that this type of nanoparticle may reach several degrees Celsius, even for pulsed lasers, possibly damaging the tissues where applied.\cite{vines_gold_2019} In addition to the advantage of having high penetration in tissues for RF, the maximum temperature rise in the cases of Figs.~\ref{fig:fig6}(c) and \ref{fig:fig6}(d) does not reach 2.6 mK. This maximum temperature rise is calculated by using a convolution solution based on the EM pulse and the Green's function of the thermal problem.\cite{egerevAcousticSignalsGenerated2009} The value was also checked by using the point-absorber approximation\cite{calassoPhotoacousticPointSource2001,egerevAcousticSignalsGenerated2009,prostPhotoacousticGenerationGold2015} since the condition \(\tau\ll{R^2}/{(4a_\text{s})}\) is valid, where \(a_\text{s}\) is the thermal diffusivity of the particle. Both calculations gave very similar results. 

As a final comment, we discuss the spatial resolution of the proposed method. It is important to remark that the spatial resolution is not specified by the electromagnetic waves since (part of) their energy is converted into acoustic energy due to photoacoustic transduction. The acoustic waves determine the resolution of the method because its application relies on enhancing or reinforcing the ultrasound signals collected by an ultrasonic transducer. As an estimate, the resolution of a single monopolar source is roughly approached by the Rayleigh limit (in 1D), as follows. Let's consider a "small-angle" approximation or a "small sensor", \cite{born_principles_2019} then the spatial resolution \(\Delta l\) can be estimated as the Rayleigh criterion, namely,
\begin{equation}
	\label{eq:spatial_resolution}
	\Delta l \sim \frac{1.22 f v_{ss}}{f_A 2R},
\end{equation}
where \(f\) is the transducer's focal distance, \(2R\) is the particle's diameter, and \(f_A\) is the acoustic frequency. For example, taking water as the surrounding medium (\(v_{ss}=1480\) m/s), \(f_A \approx 500\) MHz, and \(f = 2.5\) mm, Eq.~\ref{eq:spatial_resolution} gives a value of \(\Delta l = 902 \mu m\), which is a relatively poor resolution compared with some novel approaches \cite{yu_super-resolution_2018}. However, if we increase the particle size to \(R = 45.14 \mu m\), the resolution improves to \(\Delta l = 100 \mu m\), which matches the size of fat cells in the subcutaneous adipose tissue below \(f\) (in mm). \cite{bourdages_first-trimester_2018} Even though incrementing the particle size costs more electromagnetic energy to heat the particle and release ultrasound, diameters below \(100 \mu m\) can still be helpful for the proposed method and would be easier to fabricate. Unfortunately, a more suitable example with realistic materials has not yet been found to take total profit from the EM penetrability and avoid acoustic attenuation without losing resolution. In other words, Eq.~\ref{eq:spatial_resolution} means the resolution improves with higher acoustic frequencies but the signals' quality decreases due to ultrasound attenuation. The other option is to increase the particle size, but more EM power is required to heat the particle, and both aspects, i.e., size and power, would make the method more invasive.

All the improvements mentioned based on RF-MMs and fast RF pulses highlight the design of PA microtransducers. With the current technologies, this design may find applications in PA imaging of single cells,\cite{strohmHighResolutionUltrasound2016} \textit{in vivo} monitoring of drug delivery,\cite{abramsonIngestibleSelforientingSystem2019} and food treatment,\cite{gevekeRadioFrequencyElectric2007} among others.

\FloatBarrier

\section{Conclusion}

An unconventional photoacoustic concept was theoretically proven to be possible by simulating microparticles made of special random metamaterials mixed with soft polymers. These microcomposites have nanoelements at concentrations near the percolation limit and become conductors in the MHz range, allowing for the excitation of surface plasmons in the radiofrequency range. The soft polymer allows for the excitation of acoustic modes in the same range so that the electromagnetic and acoustic resonances  couple. This coupling  enhances  the photoacoustic pressure if the materials are carefully chosen. To fully understand the basis of the phenomenon, a solution to a general photoacoustic problem is fully developed by means of convolution products. Neither this complete solution nor its consequences were reported before. The key concepts in this  paradigm are the double-resonance and triple-resonance phenomena to enhance the ultrasound pressures and make them last longer in time. A longer signal duration translates into higher detection efficiency than in the current approaches because the pressures  also become resonant. The results illustrated here improve those obtained from the optical plasmonics with gold nanoparticles. The problems related to conventional optoacoustics, such as  thermal damage and poor tissue penetration, are avoided, in particular by using nanosecond electromagnetic pulses and MHz frequencies. 
When considering absorbing tissues as surrounding media for the particle, the intensity losses affect the effective intensity that reaches the particle, thus altering the EM absorption. This inconvenience can be sorted by increasing the intensity of the incoming wave without damaging the sample since a short pulse in the MHz region is considered in the range allowed by safety standards. As the present work is intended to show the concept of building a new type of photoacoustics combining several resonances, the problem involving embedding media with losses and superposition of several signals is considered to be beyond the scope of this paper. The only concern of changing the tissues for the phenomenon is that they may shift the electric plasmonic resonances if the real part of the permittivity of the tissue differs from that calculated for water. However, this can be easily solved by designing a new particle with an appropriate electric plasmonic resonance to function in the desired tissue. This work assumes surrounding water to mimic any embedding tissue in a general way to give a general methodology.

Remarkably, the magnetic plasmonic resonances in the particle are not disturbed by the dielectric environment that body tissues have. In other words, no magnetic response exists for any biological tissue. In this way, the double-resonance phenomenon involving magnetic and acoustic resonances remains unaltered with the tissue variation. Thus, the method involving magnetic plasmon excitation serves as a novel photoacoustic approach that may work for any tissue.

This approach suggests the possibility of real noninvasive techniques with enhanced body penetration. The proposed approach could pave the way to realize thermoacoustic imaging devices with high resolution, to be used for sensitive materials such as organic tissues, ancient artwork, or food, among other underwater applications.

\section*{Supplementary Material}

See the supplementary material for the effective medium formulas used, the thermal and mechanical properties of the materials, an analysis of skin depths in surrounding tissues and particle media, and a calculation of the PA pressure inside the particle for the case given in Figs.~\ref{fig:fig6}(c) and \ref{fig:fig6}(d).

\section*{Acknowledgments}

R.M.A.E. would like to thank Dr. Francesco De Angelis for inspiring the author to continue research in the fields of radioplasmonics and photoacoustics. Funding provided by Universidad Nacional del Centro de la Provincia de Buenos Aires is also acknowledged.

The author would also like to thank Reviewers for taking the time and effort necessary to review the manuscript. All valuable comments and suggestions were sincerely appreciated, which helped to greatly improve the quality of the manuscript.

\section*{Author declarations}
\section*{Data Availability}

This article appeared in Journal of Applied Physics, 132, 8, 083103 (2022) and may be found at (https://doi.org/10.1063/5.0086553).

The data that support the findings of this study are available
within the article and its SM.

\section*{Conflicts of Interest}
The authors have no conflicts of interest to disclose.

\end{document}